\documentclass[english,prl,showpacs,superscriptaddress,twocolumn]{revtex4}
\usepackage[T1]{fontenc}
\usepackage[latin9]{inputenc}
\usepackage{graphicx}
\usepackage{amssymb}

\makeatletter
\usepackage{esint}

\DeclareFontEncoding{LGR}{}{}

\makeatother
\begin{document}

\title{Optical wavelength conversion of quantum states with optomechanics}

\author{L. Tian}

\email{ltian@ucmerced.edu}

\affiliation{5200 North Lake Road, University of California, Merced, CA 95343,
USA}

\author{Hailin Wang}

\email{hailin@uoregon.edu}

\affiliation{Department of Physics, University of Oregon, Eugene, OR 97403, USA}

\begin{abstract}
An optomechanical interface that converts quantum states between optical fields with distinct wavelengths is proposed.  A mechanical mode couples to two optical modes via radiation pressure and mediates the quantum state mapping between the two optical modes.  A sequence of optomechanical $\pi/2$ pulses enables state-swapping between optical and mechanical states as well as the cooling of the mechanical mode.  Theoretical analysis shows that high fidelity conversion can be realized for states with small photon numbers in systems with experimentally achievable parameters.  The pulsed conversion process also makes it possible to maintain high conversion fidelity at elevated bath temperatures.
\end{abstract}
\maketitle

In quantum networks light-matter interfaces reversibly map quantum states between light and matter qubits, enabling the distribution of quantum information among distant quantum systems \cite{kimble_nature_2008}.  The quantum state mapping can be realized with electronic or spin systems, such as single atoms trapped in cavities or collective excitations in atomic or spin ensembles.  The optical wavelength of the resulting light-matter interface is set by the relevant optical transition of the electronic or spin systems. An optomechanical resonator, in which mechanical modes couple to optical modes via radiation pressure \cite{vahala_kippenberg,TeufelPRL2008,Anetsberger,Park,Eichenfield}, can also serve as a light-matter interface, in which quantum information and quantum fluctuations of an optical field can be reversibly mapped to a mechanical state \cite{Zhang, Genes, ChangKimble,Romero}.  In comparison with electronic or spin systems, the optically-active mechanical excitation can in principle couple to any of the optical modes supported by the optomechanical resonator.

\begin{figure}
\includegraphics[clip,width=7cm]{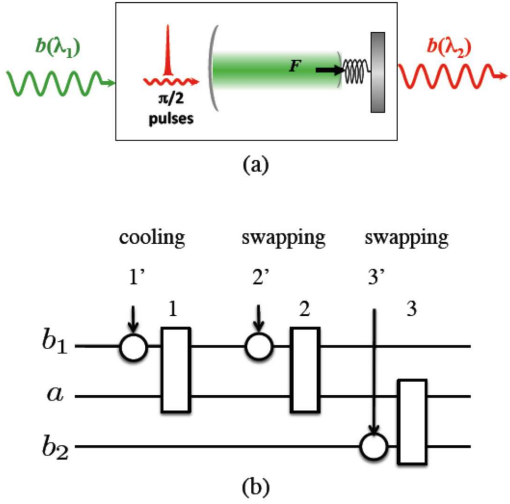}
\caption{(a) Schematic of an optomechanical system with two optical modes coupling to one mechanical mode. (b) The three-step protocol for optical wavelength conversion. The pulses $1',\,2',\,3'$ (``arrow-circle'') are initial state preparation pulses; the pulses $1,\,2,\,3$ are state swapping pulses.}
\label{fig1}
\end{figure}

Here, we propose to take advantage of this unique property of the optomechanical system to realize optical wavelength conversion of quantum states. An optical field with wavelength $\lambda_1$ can first be mapped to a mechanical state, which can then be mapped to an optical field at another wavelength $\lambda_2$, as shown schematically in Fig.~\ref{fig1}~(a). This type of interfaces, capable of optical wavelength conversion, can serve special and critical functions in a quantum network. For example, the interface can reversibly map optical fields at a given wavelength to those that are suitable for transport over long distances. The optical wavelength conversion of quantum states can also be used to interface different types of individual quantum systems such as atoms, quantum dots, or nitrogen vacancy centers in diamond, thus combining distinct advantages of the respective quantum systems.  

In this paper, we outline a scheme for the implementation of the optical wavelength conversion. A central operation of the scheme is optomechanical $\pi/2$ pulses that can induce a complete swap between optical and mechanical states.  Our scheme takes place in three steps.  In the first step (cooling), an $\pi/2$ pulse swaps the thermal state of the mechanical mode with the vacuum state of the optical mode, effectively cooling down the mechanical motion. In the second step (writing), an $\pi/2$ pulse maps an incoming optical field into a mechanical state.  In the third step (conversion), an $\pi/2$ pulse maps the mechanical state back to an optical field at another wavelength.  We have considered the optical wavelength conversion of Gaussian states as well as arbitrary states such as $(|0\rangle+|1\rangle)/\sqrt{2}$ and Schr\"{o}dinger cat states and analyzed the dependence of  the overall conversion fidelity on damping as well as the bath temperature.  We show that a conversion fidelity above the classical boundary \cite{braunsteinquantclassPRA01} can be achieved for states with a small number of photons in optomechanical resonators with experimentally realizable parameters.  In the limit that the overall duration of the conversion process is short compared with the cavity damping time, high conversion fidelity can be maintained at elevated bath temperatures.  

We consider a system consisting of two optical modes, described by annihilation operators $b_{1}$ and $b_{2}$, both coupling to a mechanical mode, described by annihilation operator $a$, via radiation pressure forces.  State swapping between the optical modes and the mechanical mode can be realized with a beam-splitter-like, linear coupling \cite{Parkins} 
\begin{equation}
H_{b}=i \epsilon_{i}( a^{\dagger} b_{i} -  b_{i}^{\dagger}a)\label{eq:Hbm}
\end{equation}
with real coupling amplitude $\epsilon_{i}$.  The dynamics of the operators can be derived as 
\begin{eqnarray}
a(t) & = & \cos(\epsilon_{i}t)a+\sin(\epsilon_{i}t)  b_{i}\nonumber \\
 b_{i}(t) & = & \cos(\epsilon_{i}t) b_{i}- sin(\epsilon_{i}t)a\label{eq:bs}
 \end{eqnarray}
 at time $t$.  For an optomechanical $\pi/2$ pulse, i.e. $\epsilon_{i}t=\pi/2$, we have $a(t)=b_{i}$ and $b_{i}(t)=-a$, which corresponds to a complete exchange of the optical state and the mechanical state except for a phase factor $-1$. 

Optomechanical couplings have the form of  $H_{G}=-\sum qG_{i}b_{i}^{\dagger}b_{i}$ where $G_{i}$'s are the coupling constants and $q=(a+a^{\dag})/\sqrt{2}$ is the displacement of mechanical vibration \cite{Law}. To engineer an effective linear coupling as Eq.(\ref{eq:Hbm}), we apply laser pulses at approximately one mechanical frequency below the relevant optical resonances (red sideband) to drive the corresponding optical modes \cite{Wilson-Rae,Tian2009,Jacobs}.  In the rotating frame of the driving, the Hamiltonian of the coupled system can be written as 
\begin{eqnarray}
H_{t} &=& \hbar\omega_{m}a^{\dag}a+\sum H_{bi}+H_{G}\label{eq:Ht} \\
H_{bi} &=& -\Delta_{i}b_{i}^{\dagger}b_{i}+(E_{i}e^{-i\omega_{0}t}b_{i}^{\dagger}+E_{i}^{*}e^{i\omega_{0}t}b_{i}) \label{eq:Hc}
\end{eqnarray}
where $\omega_{m}$ is the mechanical frequency and $H_{bi}$ is Hamiltonian of the optical modes in the rotating frame. Here, $\Delta_{i}$ is the detuning of the optical modes and $E_{i}$ is the driving amplitude.  Meanwhile, the effect of dissipation and noise on the modes can be described with a master equation approach. For the cavity modes, we use the Lindblad form $\sum_{i}(\kappa_{i}/2){\cal L}(b_{i})$ where the optical damping rate is $\kappa_{i}$ for mode $b_i$. We describe the mechanical mode with quantum Brownian motion at damping rate $\gamma_{m}$. For our system, we assume $\gamma_{m}\ll\kappa_{i},G_{i}\sqrt{\langle b_{i}^{\dag}b_{i}\rangle}$ so that effects of mechanical damping can be neglected \cite{Schwab, Blencowe}.

For the above system, the steady state average of the quadrature variables are: $q_{s}=\sum_{i}G_{i}|b_{is}|^{2}/\hbar\omega_{m}$ and $p_{s}=0$ for the mechanical mode, and 
\begin{equation}
b_{is}=\frac{-iE_{i}}{\frac{\kappa_{i}}{2}-i(\Delta_{i}+G_{i}q_{s})}\label{eq:bis}
\end{equation}
for the optical modes. By decomposing the modes in terms of the steady state average and a shifted quantum operator as $q=q_{s}+\delta q$, $p=p_{s}+\delta p$ and $b_{i}=b_{is}+\delta b_{i}$, where $\delta q$, $\delta p$, $\delta b_{i}$, and $\delta b_{i}^{\dag}$ are operators in the shifted basis,  the lowest order terms involving steady state averages cancel each other and the system can be described by an effective Hamiltonian 
\begin{eqnarray}
H_{tn} &=& \hbar\omega_{m}a^{\dag}a+\sum H_{bin}+H_{Gn} \label{eq:Htn} \\
H_{bin} &=& (-\Delta_{i}-q_{s}G_{i})\delta b_{i}^{\dag}\delta b_{i} \label{eq:Hcn} \\
H_{Gn} &=& -\sum_{i}G_{i}(b_{is}^{\star}\delta b_{i}+b_{is}\delta b_{i}^{\dagger})\delta q\label{eq:HGn}
\end{eqnarray}
where an effective linear coupling has been generated between the ``shifted'' optical and mechanical mode. With $|G_{i}b_{is}|\ll\hbar\omega_{m}$, the beam-splitter operation in Eq. (\ref{eq:Hbm}) can be achieved. 

Similar results can be derived with a quantum Langevin equation approach where the dynamics of the shifted operators is governed by linear differential equations. This approach, which has been applied to study various effects in the optomechanical systems \cite{Vitali2001}, simplifies the study of Gaussian states where the states can be described by the covariance matrix of harmonic oscillators.

The protocol for optical frequency conversion consists of three steps as illustrated in Fig.~\ref{fig1}~(b): 1. effective cooling of the mechanical mode by swapping the thermal state with the shifted vacuum state in optical mode $b_{1}$; 2. preparing and mapping the state in $b_{1}$ to the mechanical mode; 3. mapping the state of the mechanical mode  to optical mode $b_{2}$. The state swapping (transferring) pulses labelled as $1,\,2,\,3$ in Fig.~\ref{fig1}~(b)  are achieved by applying a laser pulse at approximately the red sideband,  i.e. $-\Delta_{1}-G_{1}q_{s}=\hbar \omega_{m}$ with $|G_{1}b_{1s}|\ll\hbar\omega_{m}$.  

The pulses labelled as $1',\,2',\,3'$ in Fig.~\ref{fig1}~(b) are state preparation pulses for each step. The effective linear coupling in Eq.(\ref{eq:HGn}) is between the shifted mechanical mode $\delta q$ and the shifted optical mode $\delta b_i$. In order to achieve state swapping, it is necessary to address and (or) prepare the optical (mechanical) states in the shifted vacuum.  In our scheme, we prepare the initial states through resonant excitation of cavity modes with a driving frequency $\omega_{0}=\omega_{i}$.  By adjusting the phases of the driving laser pulses, the coherent state $b_{is}$ can be seeded without (extra) coupling to the mechanical mode. After pulse $1'$, mode $b_{1}$ is prepared in the shifted vacuum (coherent state) $|b_{1s}+q_{s}\rangle$. By applying pulse $1$ for a duration of $\tau_{1}=\pi/2|G_{1}b_{1s}|$, the mechanical mode is in the shifted vacuum $|q_{s}\rangle$. After pulse $2'$, mode $b_{1}$ is prepared in the initial state $|\psi_{i}\rangle$ on top of the shifted vacuum $|b_{1s}\rangle$.  By applying pulse $2$ for a duration of $\tau_{1}=\pi/2|G_{1}b_{1s}|$, the state of the mechanical mode becomes $|\psi_{i}\rangle$ (also) on top of the shifted vacuum $|q_{s}\rangle$. After pulse $3'$, mode $b_{2}$ is prepared in the shifted vacuum $|b_{2s}\rangle$. Finally, by applying pulse $3$ satisfying $-\Delta_{2}-G_{2}q_{s}=\omega_{m}$ for a duration of $\tau_{2}=\pi/2|G_{2}b_{2s}|$, the mechanical mode recovers the shifted vacuum $|q_{si}\rangle$ and mode $b_{2}$ now is in the state $|\psi_{i}\rangle$ on top of the shifted vacuum $|b_{2s}\rangle$. 

To characterize the success of this scheme, we calculate the fidelity of the final state in mode $b_{2}$ as defined in \cite{fidelity_def}
\begin{equation}
F(\rho_{i},\rho_{f})=[\textrm{Tr}(\sqrt{\rho_{b1,i}}\rho_{b2,f}\sqrt{\rho_{b1,i}})^{1/2}]^{2}\label{eq:fid}
\end{equation}
where  $\rho_{b1,i}$ is the initial state of mode $b_1$ and $\rho_{b2,f}$ is the reduced density matrix of mode $b_{2}$ in the final state. For Gaussian states, the fidelity can be calculated directly from the reduced covariance matrix, which then simplifies the calculation for states with large photon numbers.

%
\begin{figure}
\includegraphics[width=7cm,clip]{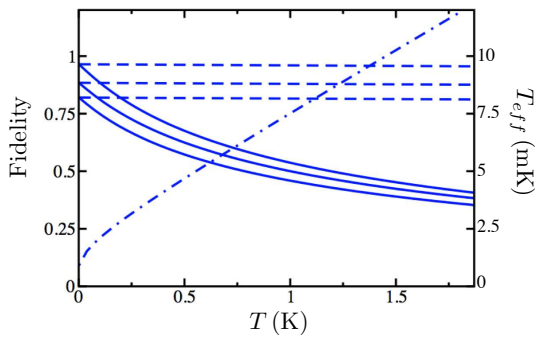}
\caption{Fidelity vs. temperature. The solid (dashed) curves are conversion fidelity at bath (effective) temperature without (with) the cooling step for Gaussian states $|\alpha=1, r_0=0\rangle$, $|\alpha=2, r_0=0\rangle$, and $|\alpha=2, r_0=0.4\rangle$ from top to bottom. The dot-dashed curve is  the effective temperature after step 1.  The parameters are $\omega_m/2\pi=100\,\textrm{MHz}$, $|G_1 b_{1s}|=10\,\textrm{MHz}$, $|G_2 b_{2s}|=7\,\textrm{MHz}$, $\kappa/2\pi=1\,\textrm{MHz}$, and $Q_m=10^4$. }
\label{fig2}
\end{figure}
The pulses in step 1 is to effectively cool the mechanical motion to the shifted vacuum $|q_{s}\rangle$. In our system, the temperature $k_{B}T$ of the thermal bath is much higher than the mechanical frequency and induces strong thermal fluctuations. In Fig.~\ref{fig2}, we plot the fidelity of mapping a state in mode $b_{1}$ to mode $b_{2}$ starting from a thermal initial state in the mechanical mode. The fidelity decreases rapidly as the bath temperature increases. With step 1, the mechanical mode can be cooled to close to the ground state.  Note that in a timescale short compared with the mechanical damping time (which is assumed to be long compared with the duration of the overall conversion process), the transient cooling process is limited only by cavity damping and by the off-resonant counter rotating terms in the effective linear coupling. We define an effective temperature $k_{b}T_{eff}$ for the mechanical state after step 1 by the relation 
\begin{equation}
\langle\delta q^{2}+\delta p^{2}\rangle=\frac{2}{e^{\hbar\omega_{m}/k_{B}T_{eff}}-1}+1\label{eq:Teff}
\end{equation}
which is plotted in Fig.~\ref{fig2}. For $T\sim2\, \textrm{K}$, we find that $T_{eff}\approx12\, \textrm{mK}$ comparable to $\hbar\omega_{m}$.  As is shown in Fig.~\ref{fig2}, there is very little loss in the conversion fidelity even for a bath temperature approaching $2\, \textrm{K}$.  

\begin{figure}
\includegraphics[width=6.4cm,clip]{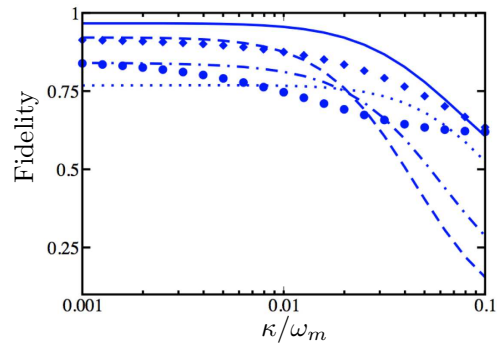}
\caption{Fidelity vs. cavity damping for various quantum states.  Solid curve: $|\alpha=1, r_0=0\rangle$; dashed curve: $|\alpha=2, r_0=0\rangle$; dot-dashed curve: $|\alpha=2, r_0=0.4\rangle$; diamonds: a superposition of Fock states $(|0\rangle+|1\rangle)/\sqrt{2}$;  and circles: Schr\"{o}dinger's cat state $\cal{N}(|\alpha\rangle+|-\alpha\rangle)$ for $\alpha=1$ with normalization factor $\cal{N}$. The dotted curve is the intermediate fidelity $F_1$ for $|\alpha=2, r_0=0.4\rangle$.  The parameters are the same as Fig.~\ref{fig2} with $T=2\,\textrm{K}$ ($T_{eff}\approx12\, \textrm{mK}$).}
\label{fig3}
\end{figure}

We further investigate the conversion fidelity by numerically simulating the process in the presence of cavity damping and finite bath temperature. In Fig.~\ref{fig3}, we plot the fidelity versus cavity damping for various Gaussian and non-Gaussian states. As the damping rate increases, the fidelity initially stays nearly flat but decreases quickly when $\kappa$ approaches $0.1|G_{i}b_{is}|$. As expected, for large $\kappa$, the leakage of photons out of the cavity significantly decreases the fidelity of state mapping.  Note that for coherent states, the classical boundary for the fidelity of state transfer is $1/2$ \cite{braunsteinquantclassPRA01}. We also observe a lower fidelity for the state $\cal{N}(|\alpha\rangle+|-\alpha\rangle)$ than that of the coherent state $|\alpha\rangle$. This is due to the difference in the small fluctuation of the $|\alpha\rangle$ component and the $|-\alpha\rangle$ component during the protocol which reduces the fidelity

The above result indicates that the key system parameter for achieving high fidelity conversion is the optical finesse relative to the effective optomechanical coupling rate (note that the optomechanical coupling constant, $G_i$, scales inversely with the size of the optical resonator). In a recent experiment, cavity damping on the order of $\kappa\sim |G_{i}b_{is}|$ has been achieved with a Fabry-Perot resonator \cite{Aspelmeyer}.  For this system, the optical finesse is approximately 14000.  A two orders of magnitude improvement in the finesse can in principle be achieved by using whispering gallery optical resonators, such as silica or CaF$_{2}$ resonators, that can feature optical finesses approaching or exceeding $10^{6}$ \cite{Braginsky,Vahala,Maliki}.  Even with such high optical finesses, the mechanical damping rate can still be small compared with the optical damping rate.  For silica resonators, a mechanical line-width of order 5 kHz (with $\omega_{m}$ near $100\, \textrm{MHz}$) can be routinely realized \cite{Anetsberger,Park}.  The mechanical Q-factor can also be improved by further reducing mechanical clamping loss.  State conversion fidelity exceeding the classical boundary can thus be realized in systems with experimentally realizable parameters.

\begin{figure}
\includegraphics[width=7cm,clip]{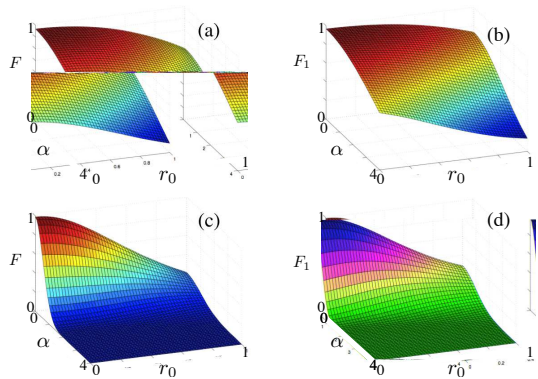}
\caption{The fidelity $F$ and intermediate fidelity $F_{1}$ vs. $\alpha$ and $r_{0}$. (a) and (b): driving at the red sideband $-\Delta_{i}-G_{i}q_{s}=\hbar \omega_{m}$. (c) and (d): driving away from the red sideband. The parameters are the same as Fig.~\ref{fig2} with $T=2\,\textrm{K}$.}
\label{fig4}
\end{figure}

We also use Gaussian states to study the dependence of fidelity on the initial state. In Fig.~\ref{fig4}, the conversion fidelity for coherent and squeezed states in the form of $|\alpha,r_0\rangle$ is plotted, where $\alpha$  is the coherent state amplitude and $r_0$ is the squeezing parameter. With small $\alpha$ and $r_{0}$, high conversion fidelity persists and decreases slowly with increasing $\alpha$ or $r_{0}$. For larger parameters, e.g. $\alpha>1$ or $r_{0}>0.4$, the fidelity deteriorates quickly with increasing parameters. For large $\alpha$, small variations of the average of the quadratures result in large deviations in the actual position of the state in the phase space, which is at the root of this sharp decrease in fidelity. Similar argument can be made for the squeezed states. Our results also show that the resonance condition $-\Delta_{i}-G_{i}q_{s}=\omega_{m}$ is crucial to achieving high fidelity. The fidelity decreases rapidly when the optical modes are driven off the anti-Stokes resonance. 

In Figs.~\ref{fig3} and \ref{fig4}, besides the overall conversion between the final state in mode $b_{2}$ and the initial state in mode $b_{1}$, we also plot the intermediate conversion fidelity $F_{1}$ between the initial state in mode $b_{1}$ and the mechanical state in mode $a$ after step 2 of the protocol. As shown in Figs.~\ref{fig3} and \ref{fig4}, in some cases, we can have $F_{2}>F_{1}$, i.e. the final optical state has higher fidelity to be in the initial state than the mechanical mode after step 2. This is because the information of the quantum state is stored in the coupled system (in the total density matrix of three modes) and is not affected by any measurement.  

In summary, we have explored the unique properties of optomechanical systems to realize optical wavelength conversion of quantum states.  The proposed implementation includes a mechanical mode couples to two optical modes via radiation pressure and exploits a sequence of optomechanical $\pi/2$ pulses for quantum state transfer and for the cooling of the mechanical mode.  We have analyzed theoretically the dependence of the overall conversion fidelity on the bath temperature, cavity damping, as well as the initial optical state.  The pulsed conversion process makes it possible to maintain high conversion fidelity at elevated bath temperatures for states with small photon numbers and in systems with experimentally achievable parameters.  An optomechanical interface that converts quantum states between optical fields with different wavelengths opens up a new and promising avenue for interfacing hybrid quantum systems and networks.  

\emph{Acknowledge.} This work is supported by the DARPA/MTO ORCHID program through a grant from AFOSR, NSF-DMR-0956064, and NSF-CCF-0916303. We also acknowledge Kurt Jacobs for discussions on cooling of a mechanical resonator via state swapping.

\end{document}